\documentclass[aps,prl,twocolumn,showpacs,superscriptaddress]{revtex4-1}

\usepackage{graphicx}
\usepackage{amsmath}
\usepackage{bm}

\newcommand{\cno}{Cu$_3$Nb$_2$O$_8$}

\begin{document}

\title{Cu$_3$Nb$_2$O$_8$: a multiferroic with chiral coupling to the crystal structure }

\author{R. D. Johnson}
\email{r.johnson1@physics.ox.ac.uk}
\affiliation{Clarendon Laboratory, Department of Physics, University of Oxford, Oxford, OX1 3PU, United Kingdom}
\affiliation{ISIS facility, Rutherford Appleton Laboratory-STFC, Chilton, Didcot, OX11 0QX, United Kingdom}
\author{Sunil Nair}
\affiliation{Clarendon Laboratory, Department of Physics, University of Oxford, Oxford, OX1 3PU, United Kingdom}
\author{L. C. Chapon}
\affiliation{ISIS facility, Rutherford Appleton Laboratory-STFC, Chilton, Didcot, OX11 0QX, United Kingdom}
\author{A. Bombardi}
\author{C.~Vecchini}
\affiliation{Diamond Light Source, Harwell Science and Innovation Campus, Didcot, OX11 0DE, United Kingdom}
\author{D. Prabhakaran}
\author{A. T. Boothroyd}
\author{P. G. Radaelli}
\affiliation{Clarendon Laboratory, Department of Physics, University of Oxford, Oxford, OX1 3PU, United Kingdom}

\date{\today}

\begin{abstract}
By combining bulk properties, neutron diffraction and non-resonant X-ray diffraction measurements, we demonstrate that the new multiferroic \cno\ becomes polar simultaneously with the appearance of generalised helicoidal magnetic ordering.  The electrical polarization is oriented perpendicularly to the common plane of rotation of the spins --- an observation that cannot be reconciled with the ``conventional'' theory developed for cycloidal multiferroics.  Our results are consistent with coupling between a macroscopic structural rotation, which is allowed in the paramagnetic group, and magnetically-induced structural chirality.

\end{abstract}


\maketitle

Magnetic multiferroics, where the electrical polarization emerges at a magnetic ordering temperature due to symmetry breaking by the magnetic structure, have been intensively studied in the past few years.  The vast majority of these new materials, such as TbMnO$_3$ \cite{kimura03,kenzelmann05}, Ni$_3$V$_2$O$_8$ \cite{lawes05} CoCr$_2$O$_4$ \cite{yamasaki06}, and MnWO$_4$ \cite{taniguchi06} are \emph{cycloidal} multiferroics, so called because their magnetic structures can be described as incommensurate circular (or elliptical) modulations with the wavevectors in the plane of rotation of the spins.  In cycloidal multiferroics, the non-collinear magnetic configuration itself is established by the competition between nearest and next-nearest neighbor interactions.  The coupling to the crystal structure occurs through the spin-orbit interaction, making it energetically favorable to develop local Dzyaloshinskii-Moriya (DM) vectors, associated with a local polarization. In simple, high-symmetry cases, the electrical polarization is perpendicular \emph{both} to the magnetic propagation vector \emph{and} to the normal to the plane of rotation of the spins, since the following formula holds \cite{mostovoy06}: ${\bm P}=\lambda \bm{k}_m \times \left(\bm{s}_1 \times \bm{s}_2\right)$, where $\bm{s}_1$ and $\bm{s}_2$ are two adjacent spins along the propagation direction $\bm{k}_m$ and $\lambda$ is a coupling constant.  In more complex, lower symmetry cases, the electrical polarization need not be perpendicular to the propagation vector, which, in turn, need not be contained in the plane of rotation of the spins (generic helicoidal structures).  However, it is a strong prediction of the cycloidal multiferroics model that if all the spins rotate in a common plane, then the electrical polarization must be strictly contained within that plane.

In this letter, we present a new multiferroic, with chemical formula \cno\ and centrosymmetric triclinic symmetry (space group $P\bar{1}$) in the paramagnetic phase. Using magnetic neutron powder diffraction, magnetic susceptibility, heat capacity, electrical polarization and non-resonant X-ray magnetic scattering measurements, we show that \cno\ orders magnetically at $T_\mathrm{N}$~$\sim$~26 K, and develops an electrical polarization below a second magnetic transition at $T_\mathrm{2}$~$\sim$~24 K.  In the polar phase below $T_\mathrm{2}$, a coplanar helicoidal magnetic structure is stabilized with propagation vector $\bm{k}_m=(0.4876, 0.2813,0.2029)$ in a general direction in reciprocal space. Strikingly, the electrical polarization in \cno\ (with a magnitude of 17.8~$\mu$Cm$^{-2}$ ) is almost exactly \emph{perpendicular} to the plane of rotation of the spins, in clear contradiction with the predictions of the cycloidal multiferroics model. We conclude that the electrical polarization in \cno\ must arise through coupling of the \emph{chiral} component of the magnetic structure with the crystal structure, rather than the ordinary \emph{polar} component --- a novel mechanism that is only allowed in a small number of paramagnetic point groups. This mechanism may also describe the magneto-electric coupling observed in other compounds reported to have ${\bm P}$ perpendicular to the spin rotation plane \cite{seki08,kimura06,arima07,kenzelmann07}.

Polycrystalline rods of \cno\ were prepared by the solid state route, with stoichiometric amounts of CuO and Nb$_2$O$_5$ being treated at 950$^\circ$~C for 36 hours. Single crystal growth was then carried out in an optical floating-zone furnace (Crystal Systems Inc.), giving high quality samples with dimensions $\sim$ 2~x~2~x~1~mm$^3$.

\begin{figure}
\includegraphics[width=8.0cm]{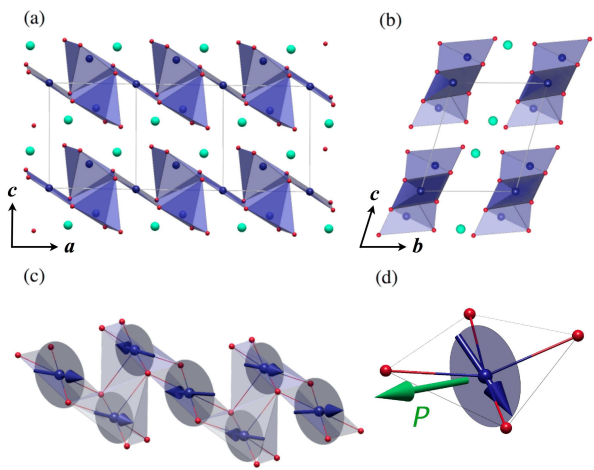}
\caption{\label{structure_fig}(Color online) (a) and (b): The crystal structure of \cno\ in the \textit{ac}- and \textit{bc}-plane, respectively. Cu$^{2+}$ ions form saw-tooth chains parallel to the \textit{a}-axis with steps of three edge sharing CuO$_4$ squares. The steps are linked via Cu2-O-Cu2 ``risers". Nb$^{5+}$ atoms separate the chains along the \textit{b}-axis. (c): The magnetic structure of the Cu$^{2+}$ chains. The envelope of the spin rotation is shown by grey circles. (d) ${\bm P}$ (green arrow) perpendicular to the spin rotation plane.}
\end{figure}

The crystal structure of \cno, not previously reported, was solved by single crystal X-ray diffraction using an Oxford Diffraction, Super Nova instrument ($\lambda=0.7107~\mathrm{\AA}$). Further details and a set of refined structural parameters are given in the Supplementary Material. Data collection and reduction was performed using the Agilent CrystAlis$^\mathrm{Pro}$ software package, and the crystal structure was refined using FullProf \cite{rodriguezcarvaja93}. A reliability factor of $R(F)$=2.94\% was achieved when fitting 691 independent reflections that met the significance criteria, $I>2.0\sigma(I)$. \cno\ was found to be isostructural to triclinic Cu$_3$NbTaO$_8$ \cite{struc_note,harneit92}. There are two distinct Cu$^{2+}$ sites in the unit cell, located at Wyckoff positions $1a$ (Cu1, on an inversion center and in square-planar coordination) and $2i$ (Cu2, in general position and square-pyramidal coordination) of space group $P\bar{1}$. The magnetic Cu$^{2+}$ sites (S=1/2) form saw-tooth chains parallel to the \textit{a}-axis with 3-atom Cu2-Cu1-Cu2 steps sharing the edges of their CuO${_4}$ squares, linked by Cu2-Cu2 ``risers'' (jumps), sharing the edges of the triangular faces of the pyramids (figures \ref{structure_fig}a and \ref{structure_fig}b). The chains are then separated along the \textit{b}-axis by a layer of non-magnetic Nb$^{5+}$ ions. The room-temperature lattice parameters of \cno\ were refined to be \textit{a}=5.1829(5)$\mathrm{\AA}$, \textit{b}=5.4857(7)$\mathrm{\AA}$, \textit{c}=6.0144(7)$\mathrm{\AA}$, $\alpha$=72.58(1)$^\circ$, $\beta$=83.421(9)$^\circ$, $\gamma$=65.71(1)$^\circ$.

\begin{figure}
\includegraphics[width=7.5cm]{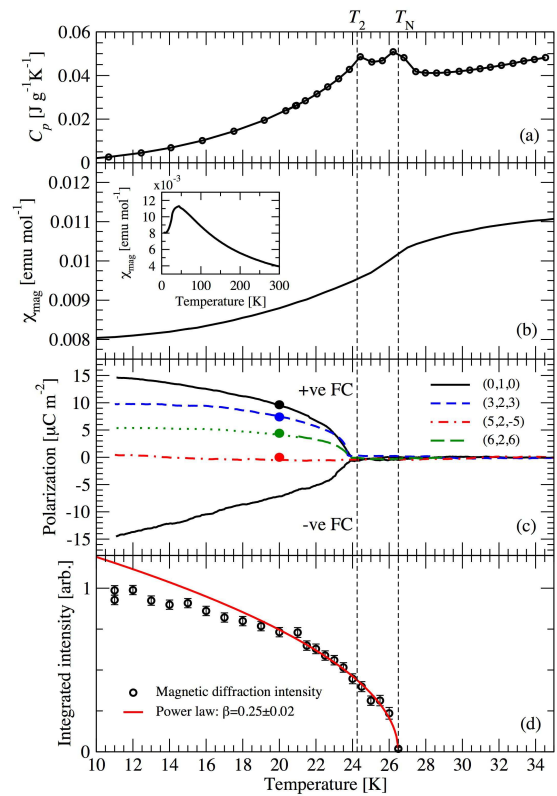}
\caption{\label{td_fig}(Color online) (a) The specific heat of \cno. (b) The magnetic susceptibility of a single crystal sample, measured in a field of 1000~Oe $\parallel$ (3,2,3). The inset shows the same measurement up to room temperature. (c) The electric polarization in three approximately orthogonal and the general (6,2,6) directions, determined through the integration of a pyroelectric current, measured at a warming rate of 1~Kmin$^{-1}$, having poled the sample with $E$~=~2~kVcm$^{-1}$. The (6,2,6) data was measured down to 21~K and extrapolated below. (d) The temperature dependence of the neutron magnetic diffraction intensity of the fundamental reflection at $d\approx 10.4~\mathrm{\AA}$. The data has been fitted with a power law.}
\end{figure}

Figure \ref{td_fig}a shows the temperature dependence of the specific heat of \cno, measured using a Quantum Design, PPMS. Two anomalies are evident, indicating phase transitions at 26.5~K and 24.2~K. Magnetic susceptibility measurements, obtained using a Quantum Design MPMS, confirmed the magnetic nature of the phase transitions (figure \ref{td_fig}b). At higher temperatures, the magnetization is dominated by a broad feature centered at 40~K, characteristic of short range magnetic correlations. A clear anomaly was observed at 26.5~K, concomitant with the first anomaly in the specific heat. Figure \ref{td_fig}c shows the electric polarization measured in four directions; (0,1,0), (3,2,3), (5,2,-5) and (6,2,6). The ferroelectric transition, labeled $T_2$, coincides with the 24.2 K anomaly in the specific heat. This behavior is similar to many of the cycloidal multiferroics \cite{kimura03,lawes05,taniguchi06}, in which an electrical polarization develops at the lower transition. The polarization vector, \textit{\textbf{P}}, was determined to be in the (1,3,2) direction with a magnitude of 17.8~$\mu$Cm$^{-2}$ at 10 K. The four polarization components were calculated at 20~K (colored dots in figure \ref{td_fig}c) and found to be in excellent agreement with the data. It is important to note that the electric polarization was 100\% switchable in electric field, as shown in figure \ref{td_fig}c for the (0,1,0) direction.

\begin{figure}
\includegraphics[width=7.5cm]{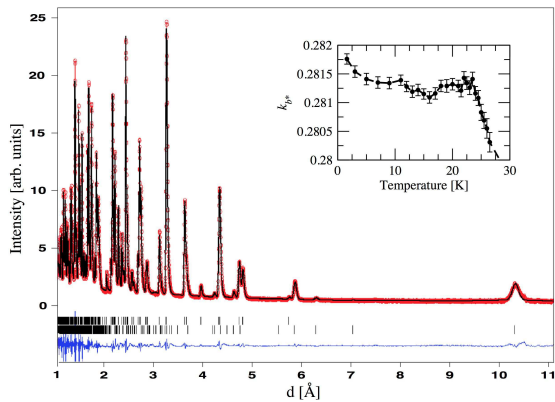}
\caption{\label{neutron_fig}(Color online) Rietveld refinement against neutron powder diffraction data of \cno\ at 1.6~K. Top and bottom tick marks indicate the nuclear and magnetic peaks, respectively. The measured and calculated profiles are shown with dots and a continuous line, respectively. A difference curve (obs.$-$calc.) is shown at the bottom. \textbf{Inset}:  temperature dependence of the $b^*$ component of $\bm{k}_m$.}
\end{figure}

A powder neutron diffraction experiment, performed on the WISH instrument \cite{chapon11}, ISIS, UK, confirmed the existence of incommensurate magnetic ordering below $T_\mathrm{N}=26.5$~K. Comparison of diffraction data measured at 1.6~K (figure \ref{neutron_fig}) and 30~K clearly showed a number of well-correlated magnetic diffraction peaks, which can be indexed using the propagation vector $\bm{k}_m~=~(0.4876, 0.2813,0.2029)$ (incommensurate in all three reciprocal space directions). Additional diffraction patterns were measured at a series of temperatures below 30~K. The magnetic propagation vector was found to be approximately constant throughout the antiferromagnetic phase, except for a pronounced change in the $b^*$ component between 24~K and 26~K (inset of figure \ref{neutron_fig}), close to the second magnetic transition at $T_\mathrm{2}$.  The strong magnetic diffraction peak at $d\approx 10.4~\mathrm{\AA}$ (the fundamental reflection where the scattering vector equals $\bm{k}_m$) was integrated to give the temperature dependence shown in figure \ref{td_fig}d, which is proportional to the square of the magnetic order parameter. Power-law fits to the temperature dependence of the magnetic diffraction intensity (continuous line in figure \ref{td_fig}d) and electrical polarization yield critical exponents of $\beta_M=0.25\pm0.02$ and $\beta_P=0.354\pm0.001$ (compatible with pseudo-proper magneto-electric coupling \cite{toledano09}).

\begin{figure}
\includegraphics[width=7.5cm]{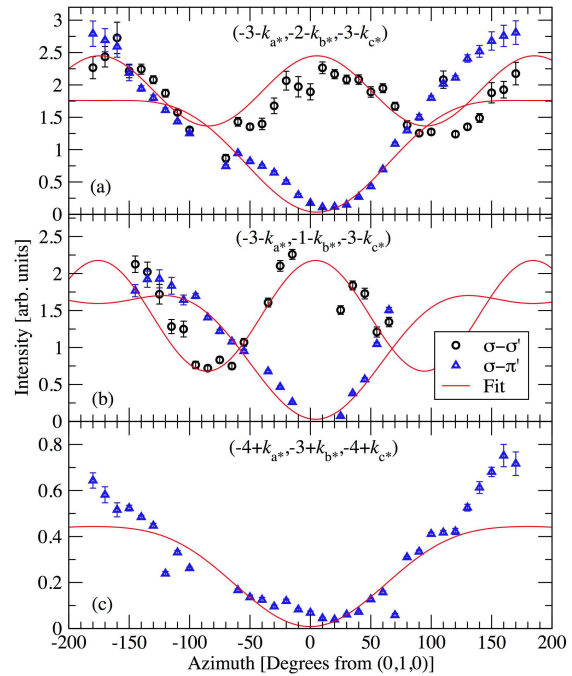}
\caption{\label{azi_fig}(Color online) Non-resonant magnetic x-ray diffraction intensity measured as a function of azimuth angle at the a) (-3-$k_{a^*}$,-2-$k_{b^*}$,-3-$k_{c^*}$), b) (-3-$k_{a^*}$,-1-$k_{b^*}$,-3-$k_{c^*}$) and c) (-4+$k_{a^*}$,-3+$k_{b^*}$,-4+$k_{c^*}$) reflections. The line shape, dependent on magnetic moment direction, has been fitted (red line) to find the spin rotation plane.}
\end{figure}

An initial fit to the neutron data with a collinear magnetic structure wrongly predicted Cu ($S=1/2$) moments to be $>~1~\mu_\mathrm{B}$.  An alternative scenario of a rotating spin order is therefore clearly favored. However, it was not possible to find a unique solution from the powder data alone. Consequently, a non-resonant magnetic X-ray diffraction (NRMXD) experiment was performed at beam line I16, Diamond Light Source, UK. This technique exploits the dependence of the NRMXD cross-section on the direction of the magnetic moment with respect to the incident and scattered directions of the light and the X-ray polarization \cite{blume85}. To minimize the fluorescent background, the incident x-ray beam was tuned to 7.835 keV, well below the copper and niobium $K$ absorption edges. Three magnetic diffraction peaks were then located. At each reflection the sample was rotated about the scattering vector (azimuthal scans) whilst measuring the diffraction intensity in both the unrotated ($\sigma-\sigma$') and rotated ($\sigma-\pi$') polarization channels, employing a graphite analyser crystal tuned to the (006) reflection. The NRMXD data are shown in figure \ref{azi_fig}. The magnetic moment directions were evaluated by fitting the azimuthal dependence of the NRMXD cross-section, which was found to be extremely sensitive to the spin rotation planes. An initial fit with independent rotation planes for the two sites (Cu1 and Cu2) converged to a single plane within the experimental error.  The normal vector of the common spin rotation plane ($\theta$ and $\phi$), and scale factors for the two sites ($F_1$ and $F_2$), then gave four free parameters to be refined.  In spherical coordinates defined with an orthonormal basis ($xyz$) such that $a\parallel x$ and $b$ is in the $xy$-plane, the normal vector of the rotation planes was found to be at $\theta=75.5(2)^\circ$ and $\phi=54.9(2)^\circ$, which corresponds approximately to the (1,2,1) reciprocal space direction. The scale factor ratio, equivalent to the ratio of the squares of the magnetic moments, was found to be $F_2/F_1=1.00(3)$. The discrepancies between calculation and data at the extremes of the $\sigma-\pi'$ azimuthal dependences are likely to be due to systematic errors inherent to the measurement. Regardless, this error in the intensity scaling does not affect the evaluation of the spin rotation plane. It may, however, lead to an imprecise determination of $F_2/F_1$, which indeed differs from that determined by the neutron diffraction measurement described in the following paragraph.

The spin rotation plane, as determined from NRMXD, has been constrained in the final Rietveld refinement of the neutron powder data, enabling us to determine the relative phases between the copper sites, the ellipticity of the spin rotation and the absolute value of the magnetic moments. The result of the refinement is plotted in figure \ref{neutron_fig}, and has reliability factors $R_p=6.46$\% and $R_{wp}=5.52$\%. The copper spins were found to describe a circular modulation (no ellipticity), with a magnitude of 0.89(2)~$\mu_\mathrm{B}$ and 0.69(1)~$\mu_\mathrm{B}$, on sites Cu1 ($1a$) and Cu2 ($2i$), respectively. The phase of the two ions occupying the $2i$ sites (inequivalent in $P1$ symmetry) were refined to be 1.03(7)$\pi$~rad and 1.05(7)$\pi$~rad relative to the ion at site $1a$. The magnetic structure, described in full in the Supplementary Material, comprises an essentially \emph{ferromagnetic} coupling within the 3-atom steps of the Cu$^{2+}$ chains, and a predominantly \emph{antiferromagnetic} coupling through the jumps, as shown in figure \ref{structure_fig}c. The propagation vector describes a generic helicoidal structure, with a fast rotation of the spins in the $b^*$ and $c^*$ directions, and a slow rotation parallel to $a^*$.   The most important result of our study was obtained by combining the results of the magnetic structure determination with those of the pyroelectric current measurements, and is illustrated in fig. \ref{structure_fig} (d).  The electrical polarization is nearly \emph{perpendicular} to the plane of rotation of the spins ($\sim 14 ^\circ$ to the vector normal to the spin rotation plane), in clear contradiction with the predictions of the cycloidal multiferroics model.  In the remainder, we will show that our result is still compatible with the inverse DM model, but calls for an altogether different interpretation.

Energy minimization in cycloidal magnets leads to the creation of local DM vectors $\bm{D}_{ij}$, associated with pairs of magnetic atoms $i$ and $j$ separated by the bond vector $\bm{r}_{ij}$.  In general, we can decompose $\bm{D}_{ij}$ in components \emph{parallel} and \emph{perpendicular} to $\bm{r}_{ij}$ as $\bm{D}_{ij}=\bm{P}_{ij} \times \bm{r}_{ij} + \sigma_{ij} \bm{r}_{ij}$, where $\bm{P}_{ij}$ is a polar vector and $\sigma_{ij}$ is a pseudoscalar (both quantities are time-reversal even).  $\bm{P}_{ij}$ can be taken to be the \emph{local} polarization, so that the macroscopic polarization is ${\bm P}=\sum_{ij} \bm{P}_{ij}$. Conversely, $\sigma=\sum_{ij} \sigma_{ij}$ is a macroscopic \emph{chirality}.  For perfect cycloidal and helical structures, $\sigma_{ij}=0$ and ${\bm P}_{ij}=0$, respectively (a cycloidal structure is \emph{not} chiral).  For a generic helicoidal structure, both $\sigma_{ij}$ and ${\bm P}_{ij}$ are present, but only the latter, which is always within the rotation plane of the spins, is considered in the cycloidal multiferroics model. The chiral term $\sigma_{ij}$ cannot in itself produce a polarization, but it can do so through coupling with the crystal structure, provided that the paramagnetic phase supports a \emph{macroscopic axial vector} ${\bm A}$, so that ${\bm P}= \gamma \sigma {\bm A}$ ($\gamma$ being a purely structural coupling constant).  Note that the electrical polarization in this case is always parallel to ${\bm A}$.  Macroscopic axial vectors represent collective rotations of one part of the structure with respect to another, and are only allowed in certain crystal classes, which we shall refer to as \emph{ferroaxial classes}.  There are seven non-polar ferroaxial classes:  $\bar{1}$, $2/m$, $\bar{3}$, $\bar{4}$, $\bar{6}$, $4/m$, $6/m$.  \cno\ belongs to the $\bar{1}$ class, the only one allowing a structural axial vector, and therefore an induced polarization, in an arbitrary direction.  Another previously unrecognized example of a ferroaxial multiferroic is RbFe(MoO$_4$)$_2$ \cite{kenzelmann07}, (space group $P \bar{3}$, crystal class $\bar{3}$) which has a proper helical magnetic structure.  In RbFe(MoO$_4$)$_2$, the electrical polarization is again perpendicular to the plane of rotation of the spins, and is constrained by symmetry along the 3-fold axis. A more intuitive way to describe this phenomenon is the following:  it is well known that the \emph{direct} DM effect produces a proper screw magnetic structure in structurally chiral MnSi \cite{kataoka81}. Conversely, a magnetic structure with a screw (helical) component will induce structural chirality by the \emph{inverse} DM effect. If the paramagnetic group is ferroaxial (i.e., supports a macroscopic rotation described by an axial vector ${\bm A}$), an electrical polarization ${\bm P}$ will develop in the direction of ${\bm A}$.  The sign of ${\bm P}$ is determined both by the direction of ${\bm A}$ and by the sign of the magnetic chirality. We note that in addition to the inverse DM effect, other more general microscopic mechanisms \cite{jia06,jia07,arima07} may apply to \cno, which are also compatible with ferroaxial coupling.

In conclusion, we have shown that in \cno\ an electrical polarization perpendicular to the plane of rotation of the spins develops below $T_2 \sim 24$~K.  We interpret this phenomenon, which cannot be explained within the framework of the cycloidal multiferroics theory, as due to the coupling between a macroscopic structural rotation (which is allowed in the paramagnetic group) and the magnetically-induced structural chirality.

\begin{acknowledgments}
RDJ would like to thank F. Fabrizi and B. H. Williams for experimental support. SN was funded by a Marie Curie IIF fellowship entitled `muti-functional-materials'.
\end{acknowledgments}

\bibliography{cno}

\begin{thebibliography}{19}%
\makeatletter
\providecommand \@ifxundefined [1]{%
 \@ifx{#1\undefined}
}%
\providecommand \@ifnum [1]{%
 \ifnum #1\expandafter \@firstoftwo
 \else \expandafter \@secondoftwo
 \fi
}%
\providecommand \@ifx [1]{%
 \ifx #1\expandafter \@firstoftwo
 \else \expandafter \@secondoftwo
 \fi
}%
\providecommand \natexlab [1]{#1}%
\providecommand \enquote  [1]{``#1''}%
\providecommand \bibnamefont  [1]{#1}%
\providecommand \bibfnamefont [1]{#1}%
\providecommand \citenamefont [1]{#1}%
\providecommand \href@noop [0]{\@secondoftwo}%
\providecommand \href [0]{\begingroup \@sanitize@url \@href}%
\providecommand \@href[1]{\@@startlink{#1}\@@href}%
\providecommand \@@href[1]{\endgroup#1\@@endlink}%
\providecommand \@sanitize@url [0]{\catcode `\\12\catcode `\$12\catcode
  `\&12\catcode `\#12\catcode `\^12\catcode `\_12\catcode `\%12\relax}%
\providecommand \@@startlink[1]{}%
\providecommand \@@endlink[0]{}%
\providecommand \url  [0]{\begingroup\@sanitize@url \@url }%
\providecommand \@url [1]{\endgroup\@href {#1}{\urlprefix }}%
\providecommand \urlprefix  [0]{URL }%
\providecommand \Eprint [0]{\href }%
\providecommand \doibase [0]{http://dx.doi.org/}%
\providecommand \selectlanguage [0]{\@gobble}%
\providecommand \bibinfo  [0]{\@secondoftwo}%
\providecommand \bibfield  [0]{\@secondoftwo}%
\providecommand \translation [1]{[#1]}%
\providecommand \BibitemOpen [0]{}%
\providecommand \bibitemStop [0]{}%
\providecommand \bibitemNoStop [0]{.\EOS\space}%
\providecommand \EOS [0]{\spacefactor3000\relax}%
\providecommand \BibitemShut  [1]{\csname bibitem#1\endcsname}%
\let\auto@bib@innerbib\@empty
\bibitem [{\citenamefont {Kimura}\ \emph {et~al.}(2003)\citenamefont {Kimura},
  \citenamefont {Goto}, \citenamefont {Shintani}, \citenamefont {Ishizaka},
  \citenamefont {Arima},\ and\ \citenamefont {Tokura}}]{kimura03}%
  \BibitemOpen
  \bibfield  {author} {\bibinfo {author} {\bibfnamefont {T.}~\bibnamefont
  {Kimura}}, \bibinfo {author} {\bibfnamefont {T.}~\bibnamefont {Goto}},
  \bibinfo {author} {\bibfnamefont {H.}~\bibnamefont {Shintani}}, \bibinfo
  {author} {\bibfnamefont {K.}~\bibnamefont {Ishizaka}}, \bibinfo {author}
  {\bibfnamefont {T.}~\bibnamefont {Arima}}, \ and\ \bibinfo {author}
  {\bibfnamefont {Y.}~\bibnamefont {Tokura}},\ }\href@noop {} {\bibfield
  {journal} {\bibinfo  {journal} {Nature}\ }\textbf {\bibinfo {volume} {426}},\
  \bibinfo {pages} {55} (\bibinfo {year} {2003})}\BibitemShut {NoStop}%
\bibitem [{\citenamefont {Kenzelmann}\ \emph {et~al.}(2005)\citenamefont
  {Kenzelmann}, \citenamefont {Harris}, \citenamefont {Jonas}, \citenamefont
  {Broholm}, \citenamefont {Schefer}, \citenamefont {Kim}, \citenamefont
  {Zhang}, \citenamefont {Cheong}, \citenamefont {Vajk},\ and\ \citenamefont
  {Lynn}}]{kenzelmann05}%
  \BibitemOpen
  \bibfield  {author} {\bibinfo {author} {\bibfnamefont {M.}~\bibnamefont
  {Kenzelmann}}, \bibinfo {author} {\bibfnamefont {A.~B.}\ \bibnamefont
  {Harris}}, \bibinfo {author} {\bibfnamefont {S.}~\bibnamefont {Jonas}},
  \bibinfo {author} {\bibfnamefont {C.}~\bibnamefont {Broholm}}, \bibinfo
  {author} {\bibfnamefont {J.}~\bibnamefont {Schefer}}, \bibinfo {author}
  {\bibfnamefont {S.~B.}\ \bibnamefont {Kim}}, \bibinfo {author} {\bibfnamefont
  {C.~L.}\ \bibnamefont {Zhang}}, \bibinfo {author} {\bibfnamefont {S.-W.}\
  \bibnamefont {Cheong}}, \bibinfo {author} {\bibfnamefont {O.~P.}\
  \bibnamefont {Vajk}}, \ and\ \bibinfo {author} {\bibfnamefont {J.~W.}\
  \bibnamefont {Lynn}},\ }\href@noop {} {\bibfield  {journal} {\bibinfo
  {journal} {Phys. Rev. Lett.}\ }\textbf {\bibinfo {volume} {95}},\ \bibinfo
  {pages} {087206} (\bibinfo {year} {2005})}\BibitemShut {NoStop}%
\bibitem [{\citenamefont {Lawes}\ \emph {et~al.}(2005)\citenamefont {Lawes},
  \citenamefont {Harris}, \citenamefont {Kimura}, \citenamefont {Rogado},
  \citenamefont {Cava}, \citenamefont {Aharony}, \citenamefont {Entin-Wohlman},
  \citenamefont {Yildrim}, \citenamefont {Kenzelmann}, \citenamefont
  {Broholm},\ and\ \citenamefont {Ramirez}}]{lawes05}%
  \BibitemOpen
  \bibfield  {author} {\bibinfo {author} {\bibfnamefont {G.}~\bibnamefont
  {Lawes}}, \bibinfo {author} {\bibfnamefont {A.~B.}\ \bibnamefont {Harris}},
  \bibinfo {author} {\bibfnamefont {T.}~\bibnamefont {Kimura}}, \bibinfo
  {author} {\bibfnamefont {N.}~\bibnamefont {Rogado}}, \bibinfo {author}
  {\bibfnamefont {R.~J.}\ \bibnamefont {Cava}}, \bibinfo {author}
  {\bibfnamefont {A.}~\bibnamefont {Aharony}}, \bibinfo {author} {\bibfnamefont
  {O.}~\bibnamefont {Entin-Wohlman}}, \bibinfo {author} {\bibfnamefont
  {T.}~\bibnamefont {Yildrim}}, \bibinfo {author} {\bibfnamefont
  {M.}~\bibnamefont {Kenzelmann}}, \bibinfo {author} {\bibfnamefont
  {C.}~\bibnamefont {Broholm}}, \ and\ \bibinfo {author} {\bibfnamefont
  {A.~P.}\ \bibnamefont {Ramirez}},\ }\href@noop {} {\bibfield  {journal}
  {\bibinfo  {journal} {Phys. Rev. Lett.}\ }\textbf {\bibinfo {volume} {95}},\
  \bibinfo {pages} {087205} (\bibinfo {year} {2005})}\BibitemShut {NoStop}%
\bibitem [{\citenamefont {Yamasaki}\ \emph {et~al.}(2006)\citenamefont
  {Yamasaki}, \citenamefont {Miyasaka}, \citenamefont {Kaneko}, \citenamefont
  {He}, \citenamefont {Arima},\ and\ \citenamefont {Tokura}}]{yamasaki06}%
  \BibitemOpen
  \bibfield  {author} {\bibinfo {author} {\bibfnamefont {Y.}~\bibnamefont
  {Yamasaki}}, \bibinfo {author} {\bibfnamefont {S.}~\bibnamefont {Miyasaka}},
  \bibinfo {author} {\bibfnamefont {Y.}~\bibnamefont {Kaneko}}, \bibinfo
  {author} {\bibfnamefont {J.-P.}\ \bibnamefont {He}}, \bibinfo {author}
  {\bibfnamefont {T.}~\bibnamefont {Arima}}, \ and\ \bibinfo {author}
  {\bibfnamefont {Y.}~\bibnamefont {Tokura}},\ }\href@noop {} {\bibfield
  {journal} {\bibinfo  {journal} {Phys. Rev. Lett.}\ }\textbf {\bibinfo
  {volume} {96}},\ \bibinfo {pages} {207204} (\bibinfo {year}
  {2006})}\BibitemShut {NoStop}%
\bibitem [{\citenamefont {Taniguchi}\ \emph {et~al.}(2006)\citenamefont
  {Taniguchi}, \citenamefont {Abe}, \citenamefont {Takenobu}, \citenamefont
  {Iwasa},\ and\ \citenamefont {Arima}}]{taniguchi06}%
  \BibitemOpen
  \bibfield  {author} {\bibinfo {author} {\bibfnamefont {K.}~\bibnamefont
  {Taniguchi}}, \bibinfo {author} {\bibfnamefont {N.}~\bibnamefont {Abe}},
  \bibinfo {author} {\bibfnamefont {T.}~\bibnamefont {Takenobu}}, \bibinfo
  {author} {\bibfnamefont {Y.}~\bibnamefont {Iwasa}}, \ and\ \bibinfo {author}
  {\bibfnamefont {T.}~\bibnamefont {Arima}},\ }\href@noop {} {\bibfield
  {journal} {\bibinfo  {journal} {Phys. Rev. Lett.}\ }\textbf {\bibinfo
  {volume} {97}},\ \bibinfo {pages} {097203} (\bibinfo {year}
  {2006})}\BibitemShut {NoStop}%
\bibitem [{\citenamefont {Mostovoy}(2006)}]{mostovoy06}%
  \BibitemOpen
  \bibfield  {author} {\bibinfo {author} {\bibfnamefont {M.}~\bibnamefont
  {Mostovoy}},\ }\href@noop {} {\bibfield  {journal} {\bibinfo  {journal}
  {Phys. Rev. Lett.}\ }\textbf {\bibinfo {volume} {96}},\ \bibinfo {pages}
  {067601} (\bibinfo {year} {2006})}\BibitemShut {NoStop}%
\bibitem [{\citenamefont {Seki}\ \emph {et~al.}(2008)\citenamefont {Seki},
  \citenamefont {Onose},\ and\ \citenamefont {Tokura}}]{seki08}%
  \BibitemOpen
  \bibfield  {author} {\bibinfo {author} {\bibfnamefont {S.}~\bibnamefont
  {Seki}}, \bibinfo {author} {\bibfnamefont {Y.}~\bibnamefont {Onose}}, \ and\
  \bibinfo {author} {\bibfnamefont {Y.}~\bibnamefont {Tokura}},\ }\href@noop {}
  {\bibfield  {journal} {\bibinfo  {journal} {Phys. Rev. Lett.}\ }\textbf
  {\bibinfo {volume} {101}},\ \bibinfo {pages} {067204} (\bibinfo {year}
  {2008})}\BibitemShut {NoStop}%
\bibitem [{\citenamefont {Kimura}\ \emph {et~al.}(2006)\citenamefont {Kimura},
  \citenamefont {Lashley},\ and\ \citenamefont {Ramirez}}]{kimura06}%
  \BibitemOpen
  \bibfield  {author} {\bibinfo {author} {\bibfnamefont {T.}~\bibnamefont
  {Kimura}}, \bibinfo {author} {\bibfnamefont {J.~C.}\ \bibnamefont {Lashley}},
  \ and\ \bibinfo {author} {\bibfnamefont {A.~P.}\ \bibnamefont {Ramirez}},\
  }\href@noop {} {\bibfield  {journal} {\bibinfo  {journal} {Phys. Rev. B}\
  }\textbf {\bibinfo {volume} {73}},\ \bibinfo {pages} {220401(R)} (\bibinfo
  {year} {2006})}\BibitemShut {NoStop}%
\bibitem [{\citenamefont {Arima}(2007)}]{arima07}%
  \BibitemOpen
  \bibfield  {author} {\bibinfo {author} {\bibfnamefont {T.}~\bibnamefont
  {Arima}},\ }\href@noop {} {\bibfield  {journal} {\bibinfo  {journal} {J.
  Phys. Soc. Jpn.}\ }\textbf {\bibinfo {volume} {76}},\ \bibinfo {pages}
  {073702} (\bibinfo {year} {2007})}\BibitemShut {NoStop}%
\bibitem [{\citenamefont {Kenzelmann}\ \emph {et~al.}(2007)\citenamefont
  {Kenzelmann}, \citenamefont {Lawes}, \citenamefont {Harris}, \citenamefont
  {Gasparovic}, \citenamefont {Broholm}, \citenamefont {Ramirez}, \citenamefont
  {Jorge}, \citenamefont {Jaime}, \citenamefont {Park}, \citenamefont {Huang},
  \citenamefont {Shapiro},\ and\ \citenamefont {Demianets}}]{kenzelmann07}%
  \BibitemOpen
  \bibfield  {author} {\bibinfo {author} {\bibfnamefont {M.}~\bibnamefont
  {Kenzelmann}}, \bibinfo {author} {\bibfnamefont {G.}~\bibnamefont {Lawes}},
  \bibinfo {author} {\bibfnamefont {A.~B.}\ \bibnamefont {Harris}}, \bibinfo
  {author} {\bibfnamefont {G.}~\bibnamefont {Gasparovic}}, \bibinfo {author}
  {\bibfnamefont {C.}~\bibnamefont {Broholm}}, \bibinfo {author} {\bibfnamefont
  {A.~P.}\ \bibnamefont {Ramirez}}, \bibinfo {author} {\bibfnamefont {G.~A.}\
  \bibnamefont {Jorge}}, \bibinfo {author} {\bibfnamefont {M.}~\bibnamefont
  {Jaime}}, \bibinfo {author} {\bibfnamefont {S.}~\bibnamefont {Park}},
  \bibinfo {author} {\bibfnamefont {Q.}~\bibnamefont {Huang}}, \bibinfo
  {author} {\bibfnamefont {A.~Y.}\ \bibnamefont {Shapiro}}, \ and\ \bibinfo
  {author} {\bibfnamefont {L.~A.}\ \bibnamefont {Demianets}},\ }\href@noop {}
  {\bibfield  {journal} {\bibinfo  {journal} {Phys. Rev. Lett.}\ }\textbf
  {\bibinfo {volume} {98}},\ \bibinfo {pages} {267205} (\bibinfo {year}
  {2007})}\BibitemShut {NoStop}%
\bibitem [{\citenamefont {Rodr{\'\i}guez-Carvajal}(1993)}]{rodriguezcarvaja93}%
  \BibitemOpen
  \bibfield  {author} {\bibinfo {author} {\bibfnamefont {J.}~\bibnamefont
  {Rodr{\'\i}guez-Carvajal}},\ }\href@noop {} {\bibfield  {journal} {\bibinfo
  {journal} {Physica B}\ }\textbf {\bibinfo {volume} {192}},\ \bibinfo {pages}
  {55} (\bibinfo {year} {1993})}\BibitemShut {NoStop}%
\bibitem [{str()}]{struc_note}%
  \BibitemOpen
  \href@noop {} {}\bibinfo {note} {\cno\ is \emph{not} isostructural to the
  well-known orthorhombic ``Kagom$\mathrm{\acute{e}}$ staircase'' multiferroic,
  Ni$_3$V$_2$O$_8$.}\BibitemShut {Stop}%
\bibitem [{\citenamefont {Harneit}\ and\ \citenamefont
  {Mueller-Buschbaum}(1992)}]{harneit92}%
  \BibitemOpen
  \bibfield  {author} {\bibinfo {author} {\bibfnamefont {O.}~\bibnamefont
  {Harneit}}\ and\ \bibinfo {author} {\bibfnamefont {H.}~\bibnamefont
  {Mueller-Buschbaum}},\ }\href@noop {} {\bibfield  {journal} {\bibinfo
  {journal} {Z. A. A. C.}\ }\textbf {\bibinfo {volume} {613}},\ \bibinfo
  {pages} {60} (\bibinfo {year} {1992})}\BibitemShut {NoStop}%
\bibitem [{\citenamefont {Chapon}\ \emph {et~al.}(2011)\citenamefont {Chapon},
  \citenamefont {Manuel}, \citenamefont {Radaelli}, \citenamefont {Benson},
  \citenamefont {Perrott}, \citenamefont {Ansell}, \citenamefont {Rhodes},
  \citenamefont {Raspino}, \citenamefont {Duxbury}, \citenamefont {Spill},\
  and\ \citenamefont {Norris}}]{chapon11}%
  \BibitemOpen
  \bibfield  {author} {\bibinfo {author} {\bibfnamefont {L.~C.}\ \bibnamefont
  {Chapon}}, \bibinfo {author} {\bibfnamefont {P.}~\bibnamefont {Manuel}},
  \bibinfo {author} {\bibfnamefont {P.~G.}\ \bibnamefont {Radaelli}}, \bibinfo
  {author} {\bibfnamefont {C.}~\bibnamefont {Benson}}, \bibinfo {author}
  {\bibfnamefont {L.}~\bibnamefont {Perrott}}, \bibinfo {author} {\bibfnamefont
  {S.}~\bibnamefont {Ansell}}, \bibinfo {author} {\bibfnamefont {N.~J.}\
  \bibnamefont {Rhodes}}, \bibinfo {author} {\bibfnamefont {D.}~\bibnamefont
  {Raspino}}, \bibinfo {author} {\bibfnamefont {D.}~\bibnamefont {Duxbury}},
  \bibinfo {author} {\bibfnamefont {E.}~\bibnamefont {Spill}}, \ and\ \bibinfo
  {author} {\bibfnamefont {J.}~\bibnamefont {Norris}},\ }\href@noop {}
  {\bibfield  {journal} {\bibinfo  {journal} {Neutron News}\ }\textbf {\bibinfo
  {volume} {22}},\ \bibinfo {pages} {22} (\bibinfo {year} {2011})}\BibitemShut
  {NoStop}%
\bibitem [{\citenamefont {Toledano}(2009)}]{toledano09}%
  \BibitemOpen
  \bibfield  {author} {\bibinfo {author} {\bibfnamefont {P.}~\bibnamefont
  {Toledano}},\ }\href@noop {} {\bibfield  {journal} {\bibinfo  {journal}
  {Phys. Rev. B}\ }\textbf {\bibinfo {volume} {79}},\ \bibinfo {pages} {094416}
  (\bibinfo {year} {2009})}\BibitemShut {NoStop}%
\bibitem [{\citenamefont {Blume}(1985)}]{blume85}%
  \BibitemOpen
  \bibfield  {author} {\bibinfo {author} {\bibfnamefont {M.}~\bibnamefont
  {Blume}},\ }\href@noop {} {\bibfield  {journal} {\bibinfo  {journal} {J.
  Appl. Phys.}\ }\textbf {\bibinfo {volume} {57}},\ \bibinfo {pages} {3615}
  (\bibinfo {year} {1985})}\BibitemShut {NoStop}%
\bibitem [{\citenamefont {Kataoka}\ and\ \citenamefont
  {Nakanishi}(1981)}]{kataoka81}%
  \BibitemOpen
  \bibfield  {author} {\bibinfo {author} {\bibfnamefont {M.}~\bibnamefont
  {Kataoka}}\ and\ \bibinfo {author} {\bibfnamefont {O.}~\bibnamefont
  {Nakanishi}},\ }\href@noop {} {\bibfield  {journal} {\bibinfo  {journal} {J.
  Phys. Soc. Jpn.}\ }\textbf {\bibinfo {volume} {50}},\ \bibinfo {pages} {3888}
  (\bibinfo {year} {1981})}\BibitemShut {NoStop}%
\bibitem [{\citenamefont {Jia}\ \emph {et~al.}(2006)\citenamefont {Jia},
  \citenamefont {Onoda}, \citenamefont {Nagaosa},\ and\ \citenamefont
  {Han}}]{jia06}%
  \BibitemOpen
  \bibfield  {author} {\bibinfo {author} {\bibfnamefont {C.}~\bibnamefont
  {Jia}}, \bibinfo {author} {\bibfnamefont {S.}~\bibnamefont {Onoda}}, \bibinfo
  {author} {\bibfnamefont {N.}~\bibnamefont {Nagaosa}}, \ and\ \bibinfo
  {author} {\bibfnamefont {J.~H.}\ \bibnamefont {Han}},\ }\href@noop {}
  {\bibfield  {journal} {\bibinfo  {journal} {Phys. Rev. B}\ }\textbf {\bibinfo
  {volume} {74}},\ \bibinfo {pages} {224444} (\bibinfo {year}
  {2006})}\BibitemShut {NoStop}%
\bibitem [{\citenamefont {Jia}\ \emph {et~al.}(2007)\citenamefont {Jia},
  \citenamefont {Onoda}, \citenamefont {Nagaosa},\ and\ \citenamefont
  {Han}}]{jia07}%
  \BibitemOpen
  \bibfield  {author} {\bibinfo {author} {\bibfnamefont {C.}~\bibnamefont
  {Jia}}, \bibinfo {author} {\bibfnamefont {S.}~\bibnamefont {Onoda}}, \bibinfo
  {author} {\bibfnamefont {N.}~\bibnamefont {Nagaosa}}, \ and\ \bibinfo
  {author} {\bibfnamefont {J.~H.}\ \bibnamefont {Han}},\ }\href@noop {}
  {\bibfield  {journal} {\bibinfo  {journal} {Phys. Rev. B}\ }\textbf {\bibinfo
  {volume} {76}},\ \bibinfo {pages} {144424} (\bibinfo {year}
  {2007})}\BibitemShut {NoStop}%
\end{thebibliography}%

\end{document}